\documentstyle[graphicx,12pt]{article}
\begin{document}

\small
\hoffset=-1truecm
\voffset=-2truecm
\title{\bf The virial relation for the Q-balls in the thermal logarithmic
potential revisited analytically}
\author {Yue Zhong \hspace {1cm}Hongbo Cheng\footnote {E-mail address:
hbcheng@ecust.edu.cn}\\
Department of Physics, East China University of Science and
Technology,\\ Shanghai 200237, China}

\date{}
\maketitle

\begin{abstract}
We study the properties of Q-balls dominated by the thermal
logarithmic potential analytically instead of estimating the
characters with only some specific values of model variables
numerically. In particular the analytical expressions for radius
and energy of this kind of Q-ball are obtained. According to these
explicit expressions we demonstrate strictly that the large
Q-balls enlarge and the small ones become smaller in the
background with lower temperature. The energy per unit charge will
not be divergent if the charge is enormous. We find that the lower
temperature will lead the energy per unit charge of Q-ball
smaller. We also prove rigorously the necessary conditions that
the model parameters should satisfy to keep the stability of the
Q-balls. When one of model parameters of Q-balls $K$ is positive,
the Q-balls will not form or survive unless the temperature is
high enough. In the case of negative $K$, the Q-balls are stable
no matter the temperature is high or low.
\end{abstract}
\vspace{5cm} \hspace{1cm} PACS number(s): 98.80.Cq, 11.27.+d,
11.10.Lm

\newpage

\noindent \textbf{I.\hspace{0.4cm}Introduction}

The Q-balls as nontopological solitons have attracted a lot of
attentions. This kind of nontopological solitons possesses a
conserved Noether charge because of a symmetry of their Lagrangian
and appears in extended localized solutions of models with certain
self-interacting complex scalar field [1, 2]. The charge of these
nontopological solitons and that their mass is smaller than the
mass of a collection of scalar fields keep them stable. In general
the Q-balls have self-interaction potential with absolute minima
and the shape of potential determines the properties of the
Q-ball. The Q-balls have been studied in many areas of physics in
order to explain the original of dark matter and the baryon
asymmetry which can not be explained by the standard model of
elementary particle physics. The Affleck-Dine mechanism produces a
scalar field condensating with baryon number while generates the
baryon asymmetry [3]. These kinds of nontopological solitons may
be considered as candidates of dark matter [4]. Recently the
scalar field configuration of the Q-ball with a step function was
considered to calculate the ratio of the Q-ball decay into the
candidates for dark matter [5]. More attentions from cosmology
were paid to the Q-balls [6-9]. In the process of expanding
universe with the sufficiently low temperature the Q-balls build
up quickly with absorbing charged particles from the outside to
result in a new kind of first-order phase transition [10]. In the
cosmological context, the existence of the Q-balls was formulated
and these kinds of Q-balls were further discussed to estimate the
net baryon number of the universe, its dark matter and the ratio
of the baryon to cold dark matter [11]. The solitosynthesis will
lead the formation of large Q-balls in the process of graduate
charge accretion if some primordial charge asymmetry and initial
seed-like Q-balls exist [12]. It was also discussed that the phase
transitions induced by solitosynthesis are possible [13]. We also
probed the nontopological solitons in de Sitter and anti de Sitter
spacetimes respectively to show the constrains from background on
the models [14]. The Q-balls can become Boson stars as flat
spacetime limits [15]. The compact Q-balls in the complex
Signum-Gordon model were also discussed [16]. The Affleck-Dine
field fragments into Q-balls which formed in the early universe
and change the scenario of Affleck-Dine baryogenesis significantly
[17].

A lot of efforts certainly have been contributed to the formation
of Q-balls. The Q-ball generates naturally in the context of
supersymmetry, in particular in Affleck-Dine mechanism for
baryogenesis [3, 18-23]. During the process the homogeneous field
as Q-ball solution begins to fluctuate and transforms into lumps.
With calculations to the full non-linear dynamics of the complex
scalar field, it was shown that the some flat directions
consisting of combination of squarks and sleptons carry the
baryonic charge in MSSM in the frame of the gravity-mediated
supersymmetry breaking scenario [22]. According to the
Affleck-Dine baryogenesis in the minimal supersymmetric standard
model with gauge-mediated supersymmetry breaking, it was found
that the Affleck-Dine field is naturally deformed into the form of
the Q-ball when the temperature is high [23]. The Q-ball formation
in the expanding universe was studied by means of 1D, 2D and 3D
lattice simulations respectively [24]. The evolution of universe
is associated with change of temperature. The thermal effects may
control the potential for Q-balls. It is necessary to research on
the Q-ball at finite temperature. The Q-ball formation in the
thermal logarithmic potential was investigated in virtue of the
lattice simulation [24]. The logarithmic potential appears during
the reheating epoch. The evolution of the universe certainly
provides the potential with the thermal correction [25]. It is
found that the Q-ball subject to the gravity-mediation potential
will transform from the thick-wall type to the thin-wall ones as
the temperature decreases and will be destroyed at last when the
temperature drops sufficiently.

It is significant to investigate the Q-ball in the thermal
logarithmic potential by means of virial theorem. This kind of
Q-balls evolving in the expanding universe could become candidates
for baryon asymmetry and dark matter. The formation and properties
of this kind of Q-ball has been estimated numerically [25, 26].
The conclusions from numerical estimation are certainly important.
It is also fundamental to find the analytical expressions for
Q-ball's charge, radius and energy to show how the Q-balls evolve
with the change of temperature in detail because the field
equations for Q-balls are nonlinear and the reliable and explicit
relation among the more model parameters as the existence of
Q-balls is difficult to be revealed by performing the burden
numerical calculation repeatedly. To our knowledge little
contribution was paid. A generalized virial relation for Q-balls
with general potential in the spacetime with arbitrary
dimensionality was obtained [27]. The analytical description
instead of a series of curves for Q-balls was derived [27]. The
analytical expressions can show clearly how the model variables
influence on the Q-balls and the fate of these kinds of
nontopological solitons. Here we will follow the procedure of Ref.
[27] to study the Q-balls controlled by the thermal logarithmic
potential. We hope to understand how this kind of Q-balls evolve
with the decreasing temperature.

In this paper we investigate the Q-ball in the thermal logarithmic
potential with virial theorem. It is difficult to solve the field
equation of Q-balls with thermal log-type potential. A virial
relation for this kind of Q-balls is found. We look for the
analytical descriptions of the radius and energy of these Q-balls
to estimate their properties in the case of large radius and small
ones respectively and to show the relation between their existence
and the expansion of the universe. We emphasize the results in the
end.

\vspace{0.8cm} \noindent \textbf{II.\hspace{0.4cm}The virial
relation for Q-balls in the thermal logarithmic potential}

We start to consider the Lagrangian density of this system as
follow,

\begin{equation}
{\mathcal{L}}=\partial_{\mu}\Phi^{+}\partial^{\mu}\Phi-V(\Phi\Phi^{+})
\end{equation}

\noindent where $\Phi=\Phi(x)$ is a complex scalar field. The
index $\mu=0,1,2,\cdot\cdot\cdot,d$ and the signature is $(+, -,
-, \cdot\cdot\cdot)$. In the Affleck-Dine scenario the homogeneous
field begins rotation with large amplitude in order to fluctuate
and transform into lumps, so the two-loop thermal effects on the
potential are crucial [28, 29]. In the case of large field values,
the potential is assumed to be

\begin{equation}
V(\Phi)=V_{T}(\Phi)+V_{m}(\Phi)
\end{equation}

\noindent where

\begin{equation}
V_{T}(\Phi)=T^{4}\ln(1+\frac{|\Phi|^{2}}{T^{2}})
\end{equation}

\begin{equation}
V_{m}(\Phi)=m_{\frac{3}{2}}^{2}|\Phi|^{2}[1+K\ln(\frac{|\Phi|^{2}}
{M^{2}})]
\end{equation}

\noindent with a global minimum at $\Phi=0$ and $T$ is
temperature. $m_{\frac{3}{2}}$ is the gravitino mass. The
parameter $K$ is negative [19, 20, 29]. Here $M$ is a
normalization scale. This potential admits the formation of
Q-balls. Under this potential Q-balls are nonperturbative
excitation about this global vacuum state carrying a net particle
number called $Q$ which is conserved. Here the energy of the
Q-ball $E_{Q}$ is smaller than $Qm_{\Phi}$ with
$m_{\Phi}^{2}=V''(0)$ because the condition will keep the Q-ball
to be stable although that is energetically preferred. The
Lagrangian of the Q-ball constrained by the thermal logarithmic
potential has a conserved $U(1)$ symmetry under the global
transformation $\Phi(x)\longrightarrow e^{i\alpha}\Phi(x)$ where
$\alpha$ is a constant. The associated conserved current density
is defined as
$j^{\mu}\equiv-i(\Phi^{+}\partial^{\mu}\Phi-\Phi\partial^{\mu}\Phi^{+})$
and the corresponding conserved charge can be given by $Q=\int
d^{d}xj^{0}$. We introduce the ansatz for field configuration with
lowest energy,

\begin{equation}
\Phi(x)=\frac{1}{\sqrt{2}}F{(\mathbf{r})}e^{i\omega t}
\end{equation}

\noindent Here the field $F(\mathbf{r})$ can be taken to be
spherically symmetry meaning $F(\mathbf{r})= F(r)$ and
$\{\mathbf{r}\}$ represents the spatial components of coordinates
of coordinates and certainly $r=|\mathbf{r}|$. The field equation
for this Q-ball can read,

\begin{equation}
(\nabla_{d}^{2}+\omega^{2})F-m_{\frac{3}{2}}^{2}KF
-\frac{2T^{4}}{F^{2}+2T^{2}}F-m_{\frac{3}{2}}^{2}(1+K\ln\frac{F^{2}}{2M^{2}})F=0
\end{equation}

\noindent From Lagrangian (1), the total energy of the system is,

\begin{equation}
E[F]=\int d^{d}x[\frac{1}{2}(\nabla_{d}F)^{2}
+\frac{1}{2}\omega^{2}F^{2}+V(F^{2})]
\end{equation}

\noindent According to Ref. [27], the virial relation is a
generalization of Derrick's theorem for Q-balls in the
$(d+1)$-dimensional spacetime can be expressed as,

\begin{equation}
d\langle
V\rangle=(2-d)\langle\frac{1}{2}(\nabla_{d}F)^{2}\rangle+\frac{d}{2}
\frac{Q^{2}}{\langle F^{2}\rangle}
\end{equation}

\noindent Since $\langle V\rangle\geq0$, the absolute lower bound
for Q-balls to be a preferred energy state is shown as,

\begin{equation}
Q^{2}\geq\frac{2(d-2)}{d}\langle
F^{2}\rangle\langle\frac{1}{2}(\nabla_{d}F)\rangle
\end{equation}

\noindent leading to

\begin{equation}
\frac{E}{Q}=\omega(1+\frac{1}{d-2+d\frac{\langle V\rangle}
{\langle\frac{1}{2}(\nabla_{d}F)^{2}\rangle}})\leq m_{\phi}
\end{equation}

\noindent In the four-dimensional spacetimes, the energy per unit
charge can be written as,

\begin{equation}
\frac{E}{Q}=\omega(1+\frac{1}{1+\frac{3\langle V\rangle}
{\langle\frac{1}{2}(\nabla F)^{2}\rangle}})
\end{equation}

\noindent where $\langle\cdot\cdot\cdot\rangle=\int\cdot\cdot\cdot
d^{3}x$ and

\begin{equation}
\langle V\rangle=\int [T^{4}\ln(1+\frac{F^{2}}{2T^{2}})
+m_{\frac{3}{2}}(1+K\ln\frac{F^{2}}{2M^{2}})F^{2}]d^{3}x
\end{equation}

\noindent The total charge of the Q-balls is,

\begin{equation}
Q=\omega\int F^{2}d^{3}x
\end{equation}

\noindent The condition (11) is fundamental to Q-balls' stability.
Because the potential has terms associated with the temperature,
the cosmic temperature decreases and so does the thermal
logarithmic potential during the process of the Universe
expansion. According to Eqs. (11) and (12) the properties
including the structure and stability of thermal log-type Q-ball
will change with time.

\vspace{0.8cm} \noindent \textbf{III.\hspace{0.4cm}The variational
approach for large Q-balls with the thermal logarithmic potential}

Here we focus on the Q-balls dominated by the thermal log-type
potential in the case of large charge and radius. First of all we
make use of the Coleman issue [2] to probe this kind of large
Q-balls. We choose the scalar field composing the Q-balls to be a
step function which is equal to be a constant denoted as $F_{c}$
within the model and vanishes outside the ball's volume $v$. The
system energy is,

\begin{equation}
E=\frac{1}{2}\frac{Q^{2}}{F_{c}^{2}v}+vV(F_{c})
\end{equation}

\noindent and
$V(F_{c})=\frac{1}{2}m_{\frac{3}{2}^{2}}(1+K\ln\frac{F_{c}^{2}}{2M^{2}})
F_{c}^{2}+T^{4}\ln(1+\frac{F_{c}^{2}}{2T^{2}})$. Having extremized
the expression (14) with respect to the volume $v$, we obtain the
minimum energy per unit charge and the condition for thermal
log-type Q-ball's stability as follow,

\begin{equation}
\frac{E_{\min}}{Q}=\frac{\sqrt{2}}{F_{c}}\sqrt{V(F_{c})}<m_{\phi}
\end{equation}

\noindent It is interesting that the minimum energy per unit
charge will be lower with decreasing temperature, which keeps the
energy density lower than the kinetic energy which is necessary
for the Q-ball to disperse. This kind of Q-balls can survive.

In order to describe the true large Q-balls with the potential
possessing the thermal logarithmic terms, we introduce the field
profile,

\begin{eqnarray}
F(r)=\{\begin{array}{cc}
  F_{c} & r<R \\
  F_{c}e^{-\alpha(r-R)} & r\geq R \\
\end{array}
\end{eqnarray}

\noindent where $\alpha$ is a variational parameter and $R$
represents a region where the field configuration keeps constant
instead of diminishing quickly. The scalar field of Q-balls can
distribute a little widely. According to the large-Q-ball ansatz
(16), the energy of model reads,

\begin{eqnarray}
E=\frac{1}{2}\frac{Q^{2}}{\langle F^{2}\rangle}
+\frac{1}{2}\langle(\nabla F)^{2}\rangle
+\langle V(F)\rangle\hspace{7cm}\nonumber\\
=\frac{1}{2}Q^{2}[\frac{4\pi}{3}F_{c}^{2}R^{3} +8\pi
F_{c}^{2}(\frac{1}{8\alpha^{3}}+\frac{R}{4\alpha^{2}}
+\frac{R^{2}}{4\alpha})]^{-1}
+4\pi\alpha^{2}F_{c}^{2}(\frac{1}{8\alpha^{3}}
+\frac{R}{4\alpha^{2}}+\frac{R^{2}}{4\alpha})\nonumber\\
+\frac{2\pi}{3}m_{\frac{3}{2}}^{2}F_{c}^{2}(^{1+K\ln\frac{F_{c}^{2}}
{2M^{2}}})R^{3}\hspace{8cm}\nonumber\\
+4\pi m_{\frac{3}{2}}^{2}F_{c}^{2}
[1+K(\ln\frac{F_{c}^{2}}{2M^{2}}+2\alpha
R)][\frac{1}{(2\alpha)^{3}}
+\frac{R}{(2\alpha)^{2}}+\frac{R^{2}}{4\alpha}]\hspace{2cm}\nonumber\\
-12\pi\pi m_{\frac{3}{2}}^{2}KF_{c}^{2}[\frac{1}{(2\alpha)^{3}}
+\frac{R}{(2\alpha)^{2}}+\frac{R^{2}}{4\alpha}+\frac{R^{3}}{6}]\hspace{4cm}\nonumber\\
+\frac{4\pi}{3}R^{3}T^{4}\ln(1+\frac{F_{c}^{2}}{2T^{2}})
\hspace{8cm}\nonumber\\ +4\pi
T^{4}\sum_{n=1}^{\infty}\frac{(-1)^{n+1}}{n}(\frac{F_{c}^{2}}
{2T^{2}})^{n}[\frac{2}{(2n\alpha)^{3}}+\frac{2R}{(2n\alpha)^{2}}
+\frac{R^{2}}{2n\alpha}]\hspace{2cm}
\end{eqnarray}

\noindent where the conserved charge is,

\begin{eqnarray}
Q=\omega\int F^{2}d^{3}x\hspace{4cm}\nonumber\\
=\frac{4\pi}{3}\omega F_{c}^{2}R^{3}+8\pi\omega F_{c}^{2}
(\frac{1}{8\alpha^{3}}+\frac{R}{4\alpha^{2}}+\frac{R^{2}}{4\alpha})
\end{eqnarray}

\noindent The above expression means that the topological charge
can take the place of the frequency $\omega$. In order to further
discuss the properties such as the stability of this kind of
Q-ball controlled by the cosmic temperature, we just leave several
dominant terms in the expressions of the energy and this
approximation is acceptable for large Q-balls. Combining the the
conserved charge and reduced energy, we have,

\begin{eqnarray}
E_{Q}\leq\frac{3Q^{2}}{8\pi F_{c}^{2}}R^{-3}
+[\frac{2\pi}{3}m_{\frac{3}{2}}^{2}F_{c}^{2}(1+K\ln\frac{F_{c}^{2}}{2M^{2}})
+\frac{4\pi}{3}T^{4}\ln(1+\frac{F_{c}^{2}}{2T^{2}})]R^{3}\hspace{1.5cm}\nonumber\\
+[\pi\alpha
F_{c}^{2}+\frac{\pi}{\alpha}m_{\frac{3}{2}}^{2}F_{c}^{2}
(1+K\ln\frac{F_{c}^{2}}{2M^{2}}-K)
+\frac{2\pi}{\alpha}T^{4}\sum_{n=1}^{\infty}\frac{(-1)^{n+1}}{n^{2}}
(\frac{F_{c}^{2}}{2T^{2}})^{n}]R^{2}
\end{eqnarray}

It should be pointed out that the variables such as radius $R$ and
coefficient $\alpha$ do not belong to the model described by
Lagrangian (1) although the energy of model depends on these
variables. We extremise the reduced energy expression (19) with
respect to $R$ and $\alpha$ respectively. We proceed performance
$\frac{\partial E}{\partial R}|_{R=R_{cl}}=0$ to find the equation
that the critical radius $R_{cl}$ of Q-balls satisfies,

\begin{equation}
3CR_{cl}^{6}+2BR_{cl}^{5}-3A=0
\end{equation}

\noindent where

\begin{equation}
A=\frac{3}{8\pi}\frac{Q^{2}}{F_{c}^{2}}
\end{equation}

\begin{equation}
B=\pi\alpha
F_{c}^{2}+\frac{\pi}{\alpha}m_{\frac{3}{2}}^{2}F_{c}^{2}
[1+K(\ln\frac{F_{c}^{2}}{2M^{2}})-K]+\frac{2\pi T^{4}}{\alpha}
\sum_{n=1}^{\infty}\frac{(-1)^{n+1}}{n^{2}}(\frac{F_{c}^{2}}{2T^{2}})^{n}
\end{equation}

\begin{equation}
C=\frac{2\pi}{3}m_{\frac{3}{2}}^{2}F_{c}^{2}(1+K\ln\frac{F_{c}^{2}}{2M^{2}})
+\frac{4\pi}{3}T^{4}\ln(1+\frac{F_{c}^{2}}{2T^{2}})
\end{equation}

\noindent The approximate solution to Eq. (20) is,

\begin{eqnarray}
R_{cl}\approx(\frac{3}{8\pi}\frac{Q^{2}}{F_{c}^{2}})^{\frac{1}{6}}
[\frac{2\pi}{3}m_{\frac{3}{2}}^{2}F_{c}^{2}(1+K\ln\frac{F_{c}^{2}}{2M^{2}})
+\frac{4\pi}{3}T^{4}\ln(1+\frac{F_{c}^{2}}{2T^{2}})]^{-\frac{1}{6}}\hspace{2.5cm}\nonumber\\
-\frac{1}{9}[\pi\alpha
F_{c}^{2}+\frac{\pi}{\alpha}m_{\frac{3}{2}}^{2}F_{c}^{2}
[1+K(\ln\frac{F_{c}^{2}}{2M^{2}})-K]+\frac{2\pi T^{4}}{\alpha}
\sum_{n=1}^{\infty}\frac{(-1)^{n+1}}{n^{2}}(\frac{F_{c}^{2}}{2T^{2}})^{n}]\nonumber\\
\times[\frac{2\pi}{3}m_{\frac{3}{2}}^{2}F_{c}^{2}(1+K\ln\frac{F_{c}^{2}}{2M^{2}})
+\frac{4\pi}{3}T^{4}\ln(1+\frac{F_{c}^{2}}{2T^{2}})]^{-1}\hspace{3cm}
\end{eqnarray}

\noindent and this solution is valid for large Q-balls at finite
temperature. It is interesting that the radius of Q-ball becomes
larger as the temperature decreases. Certainly the larger Q-balls
have larger size. We can also impose the condition $\frac{\partial
E_{Q}}{\partial\alpha}|_{\alpha=\alpha_{c}}=0$ into Eq. (19) to
obtain,

\begin{equation}
\alpha_{c}^{2}=m_{\frac{3}{2}}^{2}(1+K\ln\frac{F_{c}^{2}}{2M^{2}}-K)
+\frac{2T^{4}}{F_{c}^{2}}\sum_{n=1}^{\infty}\frac{(-1)^{n+1}}{n^{2}}
(\frac{F_{c}^{2}}{2T^{2}})^{n}
\end{equation}

\noindent According to Eq. (25),

\begin{equation}
\frac{2T^{4}}{F_{c}^{2}}\sum_{n=1}^{\infty}\frac{(-1)^{n+1}}{n^{2}}
(\frac{F_{c}^{2}}{2T^{2}})^{n}>m_{\frac{3}{2}}^{2}
(1-K\ln\frac{F_{c}^{2}}{2M^{2}}+K)
\end{equation}

\noindent where $\ln\frac{F_{c}^{2}}{2M^{2}}<0$. If $K<0$, the
inequality (26) can certainly be satisfied no matter whether the
temperature is high or low. In the case of positive parameter $K$,
the temperature can not be so low that the $\alpha_{c}$ will be
complex, then the field will oscillate outside the core to make
the Q-ball to disperse. The large Q-balls with positive parameter
$K$ will not be stable when the temperature is lower than a
critical magnitude. The inequality (26) can help us to estimate
the critical temperature for any large Q-balls with a set of
parameters. Here we prove analytically that the Q-ball should obey
the necessary conditions from Ref. [17, 24-26] instead of solving
the nonlinear differential equation numerically a lot of times
corresponding to various values of Q-ball model parameters. Our
results for positive $K$ are also consist with the relevant
conclusions in Ref. [11]. Combining Eq. (19), (24) and (25), we
find the minimum energy of large Q-balls per unit charge,

\begin{equation}
\frac{E_{Q}[F_{c}]|_{R_{c},\alpha_{c}}}{Q}=[m_{\frac{3}{2}}^{2}
+m_{\frac{3}{2}}^{2}K\ln\frac{F_{c}^{2}}{2M^{2}}+\frac{2T^{4}}{F_{c}^{2}}
\ln(1+\frac{F_{c}^{}}{2T^{2}})]^{\frac{1}{2}}
(1+\xi_{c}Q^{-\frac{1}{3}})
\end{equation}

\noindent where

\begin{eqnarray}
\xi_{c}=(\frac{9\pi}{2}F_{c}^{2})[m_{\frac{3}{2}}^{2}(1+K\ln\frac{F_{c}^{2}}{2M^{2}}-K)
+\frac{2T^{4}}{F_{c}^{2}}\sum_{n=1}^{\infty}\frac{(-1)^{n+1}}{n^{2}}
(\frac{F_{c}^{2}}{2T^{2}})^{n}]\nonumber\\
\times[m_{\frac{3}{2}}^{2}
+m_{\frac{3}{2}}^{2}K\ln\frac{F_{c}^{2}}{2M^{2}}+\frac{2T^{4}}{F_{c}^{2}}
\ln(1+\frac{F_{c}^{}}{2T^{2}})]^{-\frac{5}{6}}\hspace{2cm}
\end{eqnarray}

\noindent The asymptotic behaviour of the minimum energy of large
Q-balls with huge charge $Q$ per unit charge is

\begin{equation}
\lim_{Q\longrightarrow\infty}\frac{E_{Q}[F_{c}]|_{R_{c},\alpha_{c}}}{Q}=\sqrt{m_{\frac{3}{2}}^{2}
+m_{\frac{3}{2}}^{2}K\ln\frac{F_{c}^{2}}{2M^{2}}+\frac{2T^{4}}{F_{c}^{2}}
\ln(1+\frac{F_{c}^{}}{2T^{2}})}
\end{equation}

\noindent Here we obtain the explicit expression for the minimum
energy of large Q-balls. We can discuss the influence from the
model variables and the temperature. Having compared the Eq. (15)
with Eq. (27), we discover that the lower bound on the energy per
one particle in large Q-balls with description (16) is larger than
that of Coleman's issue because of the $\xi_{c}$-term. The minimum
energy over total charge is finite even the number of particles is
extremely large. We show the dependence of the minimum energy per
unit charge of Q-balls with thermal logarithmic potential on the
cosmic temperature for a definite charge in Fig. 1. It is clear
the minimum energy per unit charge is a decreasing function of
temperature, which means that the decreasing temperature due to
the expanding universe leads the thermal log-type Q-balls more
stable. For various magnitudes of temperature $T$, the shapes of
curves of minimum energy over our charge are similar.

\vspace{0.8cm} \noindent \textbf{IV.\hspace{0.4cm}The variational
approach for small Q-balls with the thermal logarithmic potential}

Now we pay attention to the small thermal log-type Q-balls. The
small Q-balls with radii $R\geq m_{\Phi}^{-1}$ can not be
described well with the help of thin-wall approximation. We bring
about a Gaussian ansatz in order to consider the small Q-balls
subject to thermal logarithmic potential,

\begin{equation}
F({\mathbf{r}})=F(r)=F_{c}e^{-\frac{r^{2}}{R^{2}}}
\end{equation}

\noindent Substituting the ansatz (30) into the expression (7), we
write the total energy in the case of small Q-balls at finite
temperature as follow,

\begin{eqnarray}
E=\int d^{3}x[\frac{1}{2}\omega^{2}F^{2}+\frac{1}{2}(\nabla F)^{2}
+V(F^{2})]\hspace{3cm}\hspace{2.5cm}\nonumber\\
=\frac{\sqrt{2}Q^{2}}{\pi^{\frac{3}{2}}F_{c}^{2}R^{3}}
+\frac{3\pi^{\frac{3}{2}}}{2^{\frac{5}{2}}}F_{c}^{2}R
+\frac{1}{2}(\frac{\pi}{2})^{\frac{3}{2}}m_{\frac{3}{2}}^{2}F_{c}^{2}R^{3}
[1+K\ln F_{c}^{2}-(\frac{3}{2}+\ln 2M^{2})K]\nonumber\\
+[(\frac{\pi}{2})^{\frac{3}{2}}T^{4}\sum_{n=1}^{\infty}
\frac{(-1)^{n+1}}{n^{\frac{5}{2}}}(\frac{F_{c}^{2}}{2T^{2}})^{n}]R^{3}\hspace{6cm}
\end{eqnarray}

\noindent According to Gaussian ansatz, the charge is,

\begin{equation}
Q=\omega(\frac{\pi}{2})^{\frac{3}{2}}F_{c}^{2}R^{3}
\end{equation}

\noindent replacing the frequency $\omega$. The energy for small
Q-balls controlled by thermal log-type potential just involve the
dominant terms. In order to establish the equation for the
critical radius $R_{c}$, we extremize the expression of the energy
with respect to $R$ like $\frac{\partial E_{Q}}{\partial
R}|_{R=R_{cs}}=0$, then

\begin{equation}
3cR_{cs}^{6}+bR_{cs}^{4}-3a=0
\end{equation}

\noindent where

\begin{equation}
a=\frac{1}{2}\frac{Q^{2}}{(\frac{\pi}{2})^{\frac{3}{2}}F_{c}^{2}}
\end{equation}

\begin{equation}
b=\frac{3}{2}(\frac{\pi}{2})^{\frac{3}{2}}F_{c}^{2}
\end{equation}

\begin{equation}
c=\frac{1}{2}(\frac{\pi}{2})^{\frac{3}{2}}m_{\frac{3}{2}}^{2}F_{c}^{2}
[1+\ln F_{c}^{2}-(\frac{3}{2}+\ln
2M^{2})K]+(\frac{\pi}{2})^{\frac{3}{2}}T^{4}\sum_{n=1}^{\infty}
\frac{(-1)^{n+1}}{n^{\frac{5}{2}}}(\frac{F_{c}^{2}}{2T^{2}})^{n}
\end{equation}

\noindent Similarly the acceptable approximate solution reads,

\begin{equation}
R_{cs}=[1-\frac{b}{2(9cR_{0}^{2}+2b)}]R_{0}
\end{equation}

\noindent where

\begin{equation}
R_{0}=[\frac{1}{2}\frac{Q^{2}}{(\frac{\pi}{2})^{\frac{3}{2}}F_{c}^{2}}]^{\frac{1}{6}}
\{\frac{1}{2}(\frac{\pi}{2})^{\frac{3}{2}}m_{\frac{3}{2}}^{2}F_{c}^{2}
[1+\ln F_{c}^{2}-(\frac{3}{2}+\ln
2M^{2})K]+(\frac{\pi}{2})^{\frac{3}{2}}T^{4}\sum_{n=1}^{\infty}
\frac{(-1)^{n+1}}{n^{\frac{5}{2}}}(\frac{F_{c}^{2}}{2T^{2}})^{n}\}^{-\frac{1}{6}}
\end{equation}

\noindent The critical radius of small
thermal-logarithmic-potential controlled Q-ball is shown as a
function of temperature with a definite charge $Q$. The lower
temperature will lead the small Q-balls to shrink. The minimum
energy per unit charge for small Q-balls can be expressed in terms
of the critical radius as,

\begin{equation}
\frac{E[F_{c}]|_{R=R_{cs}}}{Q}=(\frac{\tilde{a}}{c})^{\frac{1}{2}}
[2c+b(\frac{\tilde{a}}{c})^{-\frac{1}{3}}Q^{-\frac{2}{3}}-\frac{b^{2}}{18}
(\tilde{a}^{2}c)^{-\frac{1}{3}}Q^{-\frac{4}{3}}]
\end{equation}

\noindent where

\begin{equation}
\tilde{a}=\frac{1}{2(\frac{\pi}{2})^{\frac{3}{2}}F_{c}^{2}}
\end{equation}

\noindent In the limit of too much charge, The minimum energy per
unit charge for small Q-balls becomes,

\begin{equation}
\lim_{Q\longrightarrow\infty}\frac{E[F_{c}]|_{R=R_{cs}}}{Q}=
2\sqrt{\tilde{a}c}
\end{equation}

\noindent It should also be pointed out that the variable $c$ must
keep positive according to Eq. (36) and Eq. (39). When the model
parameter $K$ is negative, the variable $c$ is certainly positive.
For the Q-balls with positive $K$, only the sufficiently high
temperature can keep the variable $c$ positive. The weaker thermal
corrections must result in the nonexistence of the small Q-balls
containing positive $K$. The powers of charge $Q$ in the terms are
negative in Eq. (39), so the energy over charge will not be
divergent if the charge becomes huge. In Fig. 2, for simplicity we
also choose $F_{c}=1$ without losing generality and show that the
minimum energy per unit charge $\frac{E[\Phi_{c}]|_{R=R_{cs}}}{Q}$
also decreases in the process of universe expanding. Certainly the
energy of one particle remains smaller than the kinetic energy of
a free particle as the temperature becomes lower, which keeps the
stability of small Q-balls in the colder universe.

\vspace{0.8cm} \noindent \textbf{IV.\hspace{0.4cm}Summary}

In this work we research on the Q-balls with the thermal
logarithmic potential by means of variational estimation instead
of lattice simulation. Some numerical solutions to the nonlinear
field equation with respect to several given values of the model
parameters can not be reliable and only these numerical solutions
can not reveal the relations between the characters of the Q-balls
and their construction completely. Here we obtain the analytical
results on the Q-balls properties such as radii, energies related
to the temperature and their stability without solving the
nonlinear field equation numerically. Our analytical estimations
help us to discuss the thermal log-type Q-balls in detail and
their accuracy is acceptable. We also declare that the Q-balls
minimum energy per unit charge will not be divergent if the charge
is extremely large and the minimum energy over charge decrease to
a quantity depending on the model parameters excluding the charge
$Q$ during which the universe expands. In the colder and colder
background the energy per unit charge of Q-balls subject to the
thermal logarithmic potential becomes lower, so the Q-balls remain
stable. We also prove rigorously the necessary conditions that the
model parameters should obey for the formation and stability of
Q-balls under the influence from background temperature. The
Q-balls involving negative parameter $K$ will survive no matter
whether the temperature is high or low. The sufficiently low
temperature makes the Q-balls with positive parameter $K$ to
disappear. The expanding universe with decreasing temperature
leads the large Q-balls enlarge and small ones contract. We can
further study the related topics.

\vspace{3cm}

\noindent\textbf{Acknowledgement}

This work is supported by NSFC No. 10875043.

\newpage
\begin{figure}
\setlength{\belowcaptionskip}{10pt} \centering
  \includegraphics[width=15cm]{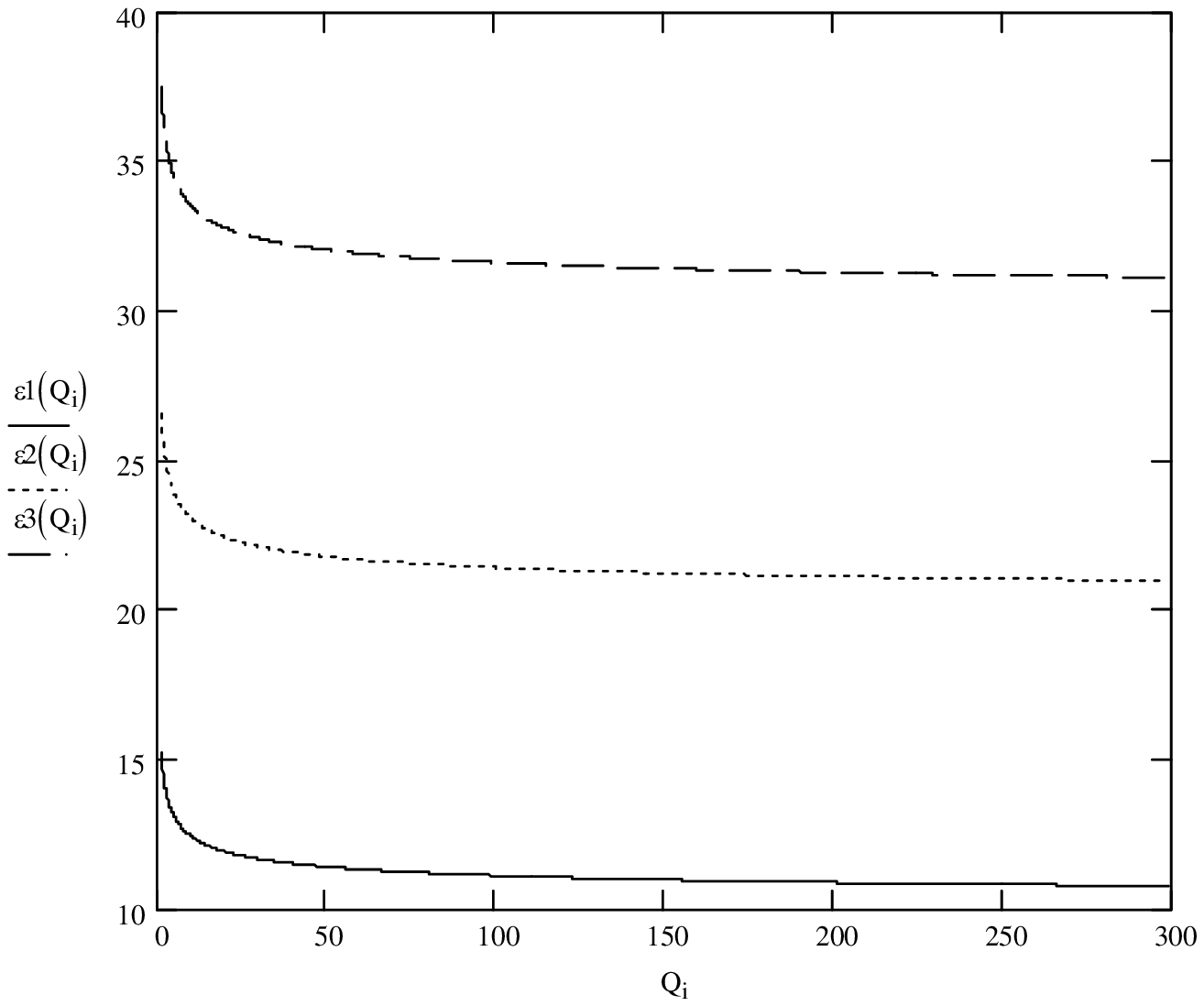}
  \caption{The solid, dot and dashed curves of the minimum
  energy per unit charge of large Q-balls in the thermal
  logarithmic potential as functions of charge $Q$
  for temperature $T=10, 20, 30$}
\end{figure}

\newpage
\begin{figure}
\setlength{\belowcaptionskip}{10pt} \centering
  \includegraphics[width=15cm]{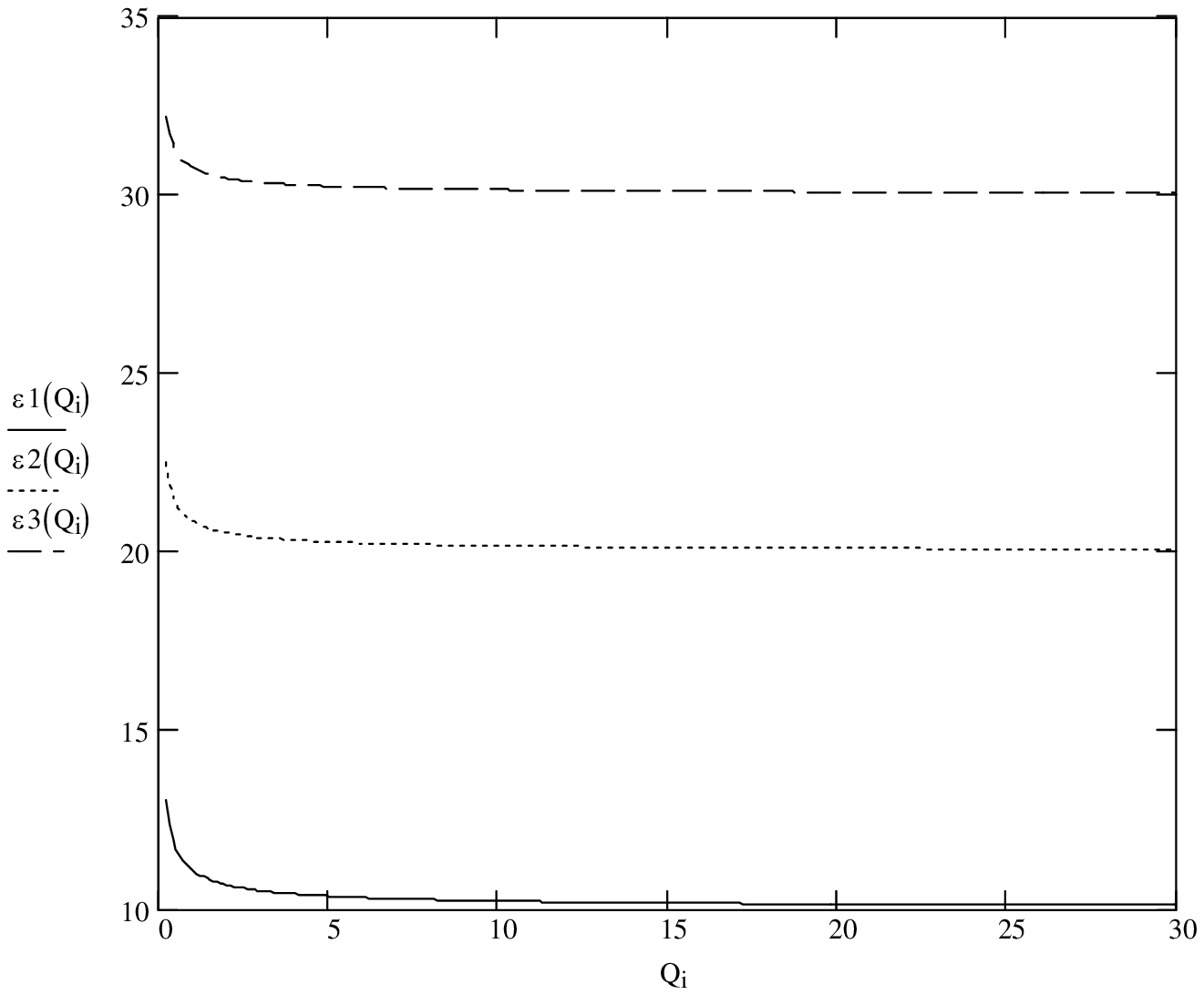}
  \caption{The solid, dot and dashed curves of the minimum
  energy per unit charge of small Q-balls in the thermal
  logarithmic potential as functions of charge $Q$
  for temperature $T=10, 20, 30$}
\end{figure}

\end{document}